%% file: paper4.tex
\documentclass[aps,pra,nobalancelastpage,superscriptaddress, twocolumn,10pt]{revtex4-2}

\usepackage{color,ifthen,amsthm,amsmath,amsxtra,amsfonts,dsfont,graphicx,bm,tikz,scalerel,wasysym,bbm,graphicx,amsthm,braket,physics,empheq,CircuitTikz}
\usepackage[colorlinks=true,linkcolor=blue, citecolor=blue, urlcolor=blue, bookmarks]{hyperref}

\newcommand{\be}{\begin{equation}}
\newcommand{\ee}{\end{equation}}
\newcommand{\bea}{\begin{eqnarray}}
\newcommand{\eea}{\end{eqnarray}}
\newcommand{\bse}{\begin{subequations}}
\newcommand{\ese}{\end{subequations}}

\theoremstyle{plain}
\newcommand{\1}{\mathbbm{1}}

\theoremstyle{plain}

\theoremstyle{plain}

\begin{document}
\title{Translation Symmetry Restoration under Random Unitary Dynamics}

\author{Katja Klobas}
\affiliation{School of Physics and Astronomy, University of Birmingham, Edgbaston, Birmingham, B15 2TT, UK}

\author{Colin Rylands}
\affiliation{SISSA and INFN Sezione di Trieste, via Bonomea 265, 34136 Trieste, Italy}

\author{Bruno Bertini}
\affiliation{School of Physics and Astronomy, University of Birmingham, Edgbaston, Birmingham, B15 2TT, UK}

\begin{abstract}

The finite parts of a large, locally interacting many-body system prepared out-of-equilibrium eventually equilibrate.  Characterising the underlying mechanisms of this process and its timescales, however, is particularly hard as it requires to decouple universal features from observable-specific ones. Recently, new insight came by studying how certain symmetries of the dynamics that are broken by the initial state are restored at the level of the reduced state of a given subsystem. This provides a high level, observable-independent probe. Until now this idea has been applied to the restoration of internal symmetries, e.g.\ $U(1)$ symmetries related to charge conservation. Here we show that that the same logic can be applied to the restoration of space-time symmetries, and hence can be used to characterise the relaxation of fully generic systems. We illustrate this idea by considering the paradigmatic example of ``generic''  many-body dynamics, i.e.\ a local random unitary circuit, where our method leads to exact results. We show that the restoration of translation symmetry in these systems only happens on time-scales proportional to the subsystem's volume. In fact, for large enough subsystems the time of symmetry restoration becomes initial-state independent (as long as the latter breaks the symmetry at time zero) and coincides with the thermalisation time. For intermediate subsystems, however, one can observe the so-called ``quantum Mpemba effect'', where the state of the system restores a symmetry faster if it is initially more asymmetric. We provide the first exact characterisation of this effect in a non-integrable system.

\end{abstract}

\maketitle

One of the most basic, yet lucrative tasks a scientist can perform when analysing a physical system is to identify its symmetries and their breaking mechanisms. This can immediately return a wealth of information regarding the possible processes that can occur within the system, the nature of the excitations, and its potential phases. When applied to quantum many-body systems at equilibrium this analysis has provided enormous insight into the allowed states of quantum matter and facilitated breakthroughs in all areas of theoretical physics from the standard model to superconductivity~\cite{weinberg1996quantum2, altland2010condensed}. More recently this perspective has been successfully applied also in out-of-equilibrium settings~\cite{vaccaro2008tradeoff, gour2009measuring, marvian2014extending, laflorencie2014spin, goldstein2018symmetry, xavier2018equipartition, bonsignori2019symmetry, denardis2022correlation, doyon2022hydrodynamic, doyon2023ballistic, gopalakrishnan2024distinct, bertini2023nonequilibrium, murciano2020entanglement, ares2022entanglement, bertini2024dynamics,khemani2018operator,rakovszky2018diffusive,friedman2019spectral,agrawal2022entanglement,barratt2022field}. 

In the latter context a simple, yet instructive, question concerns symmetry restoration from non-symmetric initial states. Namely, one considers an initial state that breaks a symmetry of the evolution operator and tracks the symmetry restoration within a subsystem. Ref.~\cite{ares2022entanglement} showed that this process can be characterised in a direct, observable-independent way by measuring the distance between the evolution of the state and its symmetrised counterpart. This also reveals an unexpected phenomenon that has been dubbed  the \emph{quantum Mpemba effect} which occurs when a symmetry is restored more rapidly when it is broken more by the initial state~\cite{ares2022entanglement}. This genuine quantum effect owes its name to its similarity with the ``classical'' Mpemba effect, arising when hot water freezes faster than cold water~\cite{mpemba1969cool}, and has been observed in a range of different scenarios~\cite{ares2022entanglement,ares2023lack,murciano2024entanglement,khor2024confinement,ferro2023nonequilibrium,capizzi2023entanglement,capizzi2024universal,bertini2024dynamics,rylands2024microscopic,chen2024renyi,fossati2024entanglement,caceffo2024entangled,ares2024entanglement,liu2024symmetry,yamashika2024entanglement,ares2025quantum,chalas2024multiple,turkeshi2024quantum}, including trapped ion quantum simulators~\cite{joshi2024observing}. 

Up to now, however, this observable-independent characterisation of the symmetry restoration has only been performed for systems with certain special internal symmetries, mostly $U(1)$ or its discrete subgroups. Here we extend this approach to the case of arbitrary symmetries including \emph{spatial} ones.  As a relevant application we then use it to characterise \emph{exactly} the restoration of translation symmetry in random unitary circuits, where the evolution is implemented by local unitary matrices that are independently drawn from the Haar distribution at each space-time point~\cite{fisher2023random}. These systems possess no internal symmetries, relax locally to the infinite temperature state, and provide an efficient scrambling of quantum information making them the paradigmatic example of quantum chaotic system~\cite{nahum2017quantum, nahum2018operator, vonKeyserlingk2018operator, khemani2018operator, zhou2019emergent, zhou2020entanglement, bertini2020scrambling}. Their averaged dynamics, however, does exhibit space translation symmetry and one can ask how the latter is restored when broken by the initial state. 

In this setting, the intuition based on the entanglement dynamics suggests almost immediate restoration of the translational-symmetry. Indeed, when quenching from product states, R\'enyi entropies lose any initial-state dependence upon averaging~\cite{nahum2018operator}, while for more complicated initial state the state-dependence is subleading~\footnote{This follows from our results in Ref.~\cite{Note11}.}. Given the fact that R\'enyi entropies characterise the full spectrum of the density matrix, these results seem to imply that the latter should lose its initial-state dependence very quickly. In contrast, however,
we  find that the symmetry is not fully restored for any finite time --- signaling that even in these random systems, subsystems preserve memory of the initial state for long times. If we ask for $\epsilon$-approximate restoration, i.e., for symmetry restoration up to a correction $\epsilon$, we find that, for large enough subsystems, the restoration time scales linearly with the subsystem size,  and is initial-state independent. Despite this we observe occurrences of the quantum Mpemba effect for subsystems of intermediate sizes, and we provide its first exact characterisation in a non-integrable system.

More specifically, we consider a quantum many-body system with Hilbert space $\mathcal H$ and where the time-evolution operator $\mathbb U$ has a symmetry group $\mathcal G$ admitting a unitary representation $\mathcal G\ni g\mapsto \Pi_g \in {\rm End}{(\mathcal H)}$~\footnote{Symmetries associated to anti-Unitary representations can also be straightforwardly accommodated.}. Then we focus on a quantum quench protocol preparing the system in a state $\ket{\Psi}$ that is neither an eigenstate of $\mathbb U$, nor of the symmetry group. To quantify the amount of symmetry breaking caused by the initial state, and study its restoration in time, we measure the distance between the reduced density matrix of a subsystem $A$, $\rho_A(t)= \tr_{\bar A}[ \ketbra{\Psi(t)}{\Psi(t)} ]$, and the reduced \emph{symmetrised} state 
\begin{align}
  \label{eq:symstate}
  \bar{\rho}_A(t)&=\frac{1}{|\mathcal G|}\sum_{g\in\mathcal G} \rho^{(g)}_{A}(t),\\
  \label{eq:symstate2}
  \rho^{(g)}_{A}(t) 
  &= \tr_{\bar A}[\mathbb U^t \Pi_g \ketbra{\Psi(0)}{\Psi(0)} \Pi_g^{-1} \mathbb U^{-t}],
\end{align}
where $|\mathcal G|$ is the order of the group~\footnote{In writing this expression we assumed that $\mathcal G$ is a finite group. Our treatment, with obvious modifications, also applies to the case of $\mathcal G$ being a Lie group.}. Note that when the representation of the symmetry group factorises between $A$ and its complement, i.e., $\Pi_g =  \Pi_{g,A} \otimes  \Pi_{g, \bar A}$ for all $g$, the symmetrised state can be written as a completely-positive trace-preserving map on $\rho_A(t)$, i.e.
\be
\label{eq:simprhobar}
\bar{\rho}_A(t)=\frac{1}{|\mathcal G|}\sum_{g\in\mathcal G} 
\Pi_{g, A} \rho_A(t) \Pi_{g, A}^{-1}.
\ee 
A notable example where this factorised form holds is when $\mathcal G$ is generated by a charge with density supported on a single site, e.g.,  particle number. So far, all the studies of subsystem symmetry restoration after quenches focussed on this special case~\cite{ares2022entanglement,ares2023lack,murciano2024entanglement,khor2024confinement,ferro2023nonequilibrium,capizzi2023entanglement,capizzi2024universal,bertini2024dynamics,rylands2024microscopic,chen2024renyi,fossati2024entanglement,caceffo2024entangled,ares2024entanglement,liu2024symmetry,yamashika2024entanglement, joshi2024observing,ares2025quantum,chalas2024multiple,turkeshi2024quantum} while here we do not impose this restriction. 
 
To quantify the distance between $\rho_A(t)$ and $\bar \rho_A(t)$ we compute the normalised Frobenius distance 
\be
\label{eq:Frobdistnorm}
\Delta_2(t) = \frac{\|\rho_A(t)-\bar \rho_A(t)\|_2}{\|\rho_A(t)\|_2},
\ee
where $\|O\|_2=\sqrt{\tr[OO^\dag]}$ denotes the Frobenius norm. By definition, this quantity is always positive and equal to zero only if $\rho_A(t)= \bar \rho_A(t)$. We have used \eqref{eq:Frobdistnorm}, instead of the quantum relative entropy as in Ref.~\cite{ares2022entanglement} and subsequent works, because, whenever the simplified form~\eqref{eq:simprhobar} does not apply, the latter becomes much harder to access analytically. Essentially this is because to compute it one needs to introduce a replica trick~\cite{ruggiero2017relative} that breaks the non-negativity property.

Our goal is to use \eqref{eq:Frobdistnorm} to characterise translation symmetry restoration in random unitary circuits. In particular, we consider a one dimensional chain of $2L$ qudits with local Hilbert space $\mathcal H = \mathbb{C}^q$, $2\le q\in\mathbb{Z}$.  The coordinates of the qudits are indicated by $x\in \mathbb{Z}/2$ and we assume periodic boundary conditions, $x+L=x$.  The dynamics of the system are generated by a brickwork-like, random quantum circuit such that the state of the system after $t+1$ time steps is given by 
\be\label{eq:time_evo_op}
\ket{\Psi(t+1)}=\mathbb{U}(t+1)\ket{\Psi(t)},
\ee
and where the time-evolution operator is 
\be 
 \mathbb{U}(t+1)=
\smashoperator{\bigotimes_{x\in\mathbb{Z}_L+\frac{1}{2}}} U_{x,x+\frac{1}{2}}(x,t)
\smashoperator{\bigotimes_{x'\in\mathbb{Z}_L}} U_{x'\!\!,x'+\frac{1}{2}}(x',t).
\ee
Here $U_{x,x+{1}/{2}}(x,t)$ are operators acting on the whole chain, but acts non-trivially --- as the matrix $U(x,t)$ --- only at sites $x,x+{1}/{2}$, and as the identity everywhere else. The $q^2\times q^2$ matrices $U(x,t)$ are known as the \emph{local gates}. In our system we take them to be chosen independently from the Haar random ensemble.

For each different realisation of the $2Lt$ local gates the system exhibits no conserved quantities and has no symmetry.  Upon averaging over different realisations however,  the dynamics acquires a $\mathbb Z_{L}$ symmetry under discrete translations. In particular
\begin{align}
\expval{\Pi \mathbb{U}(t)\Pi^\dag} =\expval{\mathbb{U}(t)},
\end{align}
where we have denoted the ensemble average by $\left<\cdot\right>$ and introduced the two-site shift operator $\Pi$ which translates the entire system by two sites to the right ($\Pi^L=\1$). Namely, for an operator $\mathcal{O}_{x}$ acting on site $x$ we have
$
\Pi\mathcal{O}_{x} \Pi^\dag=\mathcal{O}_{x+1}.
$
This emergent symmetry is manifested in the long time limit of the state of the subsystem $A$, which we take to be a block of $2\ell$ qubits. It is easy to see that after averaging over different realisations the latter is given by the infinite temperature state $\1/q^{2\ell}$. Therefore, the long time steady state is invariant under the action of two-site shifts and $\lim_{t\to\infty} \expval{ \rho_A(t)}=\1/q^{2\ell}=\lim_{t\to\infty} \expval{ \bar\rho_A(t)}$. 

 \begin{figure}
\includegraphics[width=\columnwidth, trim= 20 0 80 0, clip]{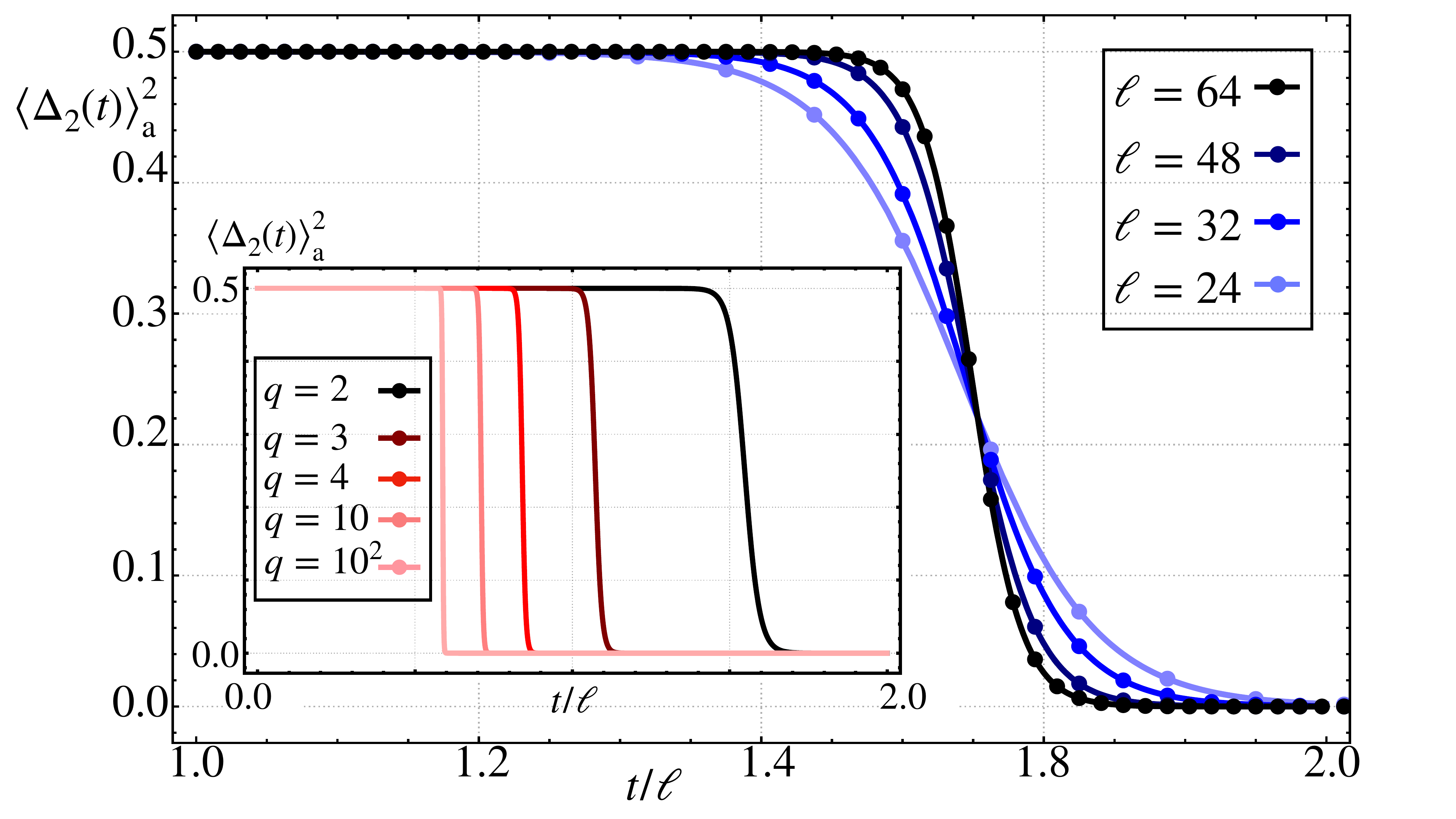}
  \caption{\label{fig:twistedstate}
 Dynamical restoration of $2$-site translational symmetry in random unitary circuits.  We plot the  squared Frobenious distance,  $\expval{\Delta_2(t)}^2_{\rm a}$,  between $\rho_A(t)$ and the symmetrised state $\bar \rho_A(t)$,   as function of $t/\ell$ for different subsystem sizes $|A|=2\ell$.  The initial state is parameterized by \eqref{eq:initialstate} with $c=1$.  Different lines correspond to increasing $\ell=24,32,48,64$, lighter to darker. The symbols are the exact result while the solid lines are the result of~\eqref{eq:approximatesolution}. In the inset we plot the same for $\ell=48$ and increasing local Hilbert space dimension $q=2,3,4,10,100$, darker to lighter.}
\end{figure}

We wish to study the emergence of this spatial symmetry for a system prepared in a state that explicitly breaks it. We focus on states which are formed from products over two sites of the form
\be \label{eq:initial_state}
\ket{\Psi(0)}=\bigotimes_{x=1}^L\ket{m_x},\,\, 
\ket{m_x} \!=\!\!\sum_{i,j=1}^{q}[m_x]_{ij}\ket{i}_x\!\otimes\ket{j}_{x+\frac{1}{2}}\!, 
\ee
where $\ket{i}_x,i=1,\dots q$ span the set of states of the qudit at site $x$ and $\{m_x\}$ is a set of $q\times q$ matrices characterizing the initial state which are taken to be properly normalised $\tr[(m_x)^\dag m_x]=1$. In addition, we impose that the states are translationally invariant only over shifts of $2\nu$ sites with ${\nu>1}$~\footnote{Necessarily, we require that $L/\nu\in \mathbb{N}$.}.
{Note that under translational invariant dynamics the operator $\Pi_g$ in Eq.~\eqref{eq:symstate2} commutes with the time-evolution operator. Therefore, one can either  apply it to the initial state or to the state at time $t$. This is not the case for random unitary circuits, as the translational symmetry emerges only after ensemble averaging.}

Since by construction the Frobenius distance is non negative and we have ${\expval{\Delta_2(0)}>0}$ and ${\expval{\Delta_2(\infty)}=0}$, we can use $\expval{\Delta_2(t)}$ to track the emergence of the spatial symmetry. In fact, to aid our analytic calculations we consider 
\be
\label{eq:Frobapp}
\expval{\Delta_2(t)} \approx \expval{\Delta_2(t)}_{\rm a}  \equiv \sqrt{\frac{\expval{\|\rho_A(t)-\bar \rho_A(t)\|_2^2}}{\expval{\|\rho_A(t)\|_2^2}}},
\ee
where we have taken the average inside the square root and separately averaged numerator and denominator. Previous work has established that these manipulations do not change appreciably the result due to small circuit to circuit fluctuations~\cite{nahum2018operator,vonKeyserlingk2018operator,zhou2019emergent,ares2024entanglement,turkeshi2024quantum}. Note that also the quantity on the r.h.s.\ of \eqref{eq:Frobapp} is non-negative and only vanishes for $\rho_A(t)=\bar \rho_A(t)$, therefore it provides a good measure of the restoration of spatial symmetry.

Let us now illustrate the main steps in the calculation of  $\expval{\Delta_2(t>0)}_{\rm a}$ and discuss its central features. For pedagogical reasons we restrict ourselves to case $\nu=2$, with the general case discussed in \cite{Note11}. \footnotetext[11]{See the Supplemental Material for details on the solution to the recurrence relation for the quantity $D_{y}$, and the discussion of the $\nu>2$ case.} As in previous studies of subsystem symmetry restoration~\cite{ares2022entanglement,ares2023lack,murciano2024entanglement,khor2024confinement,ferro2023nonequilibrium,capizzi2023entanglement,capizzi2024universal,bertini2024dynamics,rylands2024microscopic,chen2024renyi,fossati2024entanglement,caceffo2024entangled,ares2024entanglement,liu2024symmetry,yamashika2024entanglement, joshi2024observing,ares2025quantum,chalas2024multiple}, the central object to study is the averaged overlap between the states in the linear combination~\eqref{eq:symstate}. Specifically, setting ${\rho^{(0)}_{A}(t)=\rho_A(t)}$ and $\rho_{A}^{(1)}(t)=\rho^{(\Pi)}_{A}(t)$ we express~\eqref{eq:Frobapp} as 
\be
\!\!\!\!\!\expval{\Delta_2(t)}_{\rm a}^2 \!=\! \frac{1}{4}\!+\! \frac{\expval{\tr[\rho^{(1)}_{A}(t)^2]}\!-\!2 \expval{\tr[\rho^{(0)}_{A}(t)\rho^{(1)}_{A}(t)]}}{4 \expval{{\tr[{\rho_{A}^{(0)}(t)^2}]}}}. 
\label{eq:delta2overlap}
\ee
Noting that $\tr\smash{[\rho^{(0)}_{A}(t)\rho^{(1)}_{A}(t)]}\geq 0$, because $\rho^{(0)}_{A}(t)$ and $\rho^{(1)}_{A}(t)$ are both positive semidefinite, we have that the Frobenius distance attains its maximal value (of $1/2$) when the two states are orthogonal. Conversely it is zero only if $\rho^{(0)}_{A}(t)=\rho^{(1)}_{A}(t)$. 

Using the special form of the Haar-averaged local gate one can show that averaged overlaps fulfil a simple recurrence relation, which can be solved in closed form. To see this it is convenient to represent the overlaps diagrammatically using the standard graphical notation of tensor networks~\cite{cirac2021matrix}. Specifically, we represent the averaged, folded, local gate as
\be
  W=\left<\right.\!\left(U(x,t)\otimes_r U^{\star}(x,t)\right)^{\otimes_{\rm{r}} 2}\!\left.\right>=
  \begin{tikzpicture}[baseline={([yshift=-0.6ex]current bounding box.center)},scale=0.5]
   \prop{0}{0}{FcolU}{}
 \end{tikzpicture}~,
\ee
where we have denoted by $\otimes_{\rm{r}}$ the tensor product over different replicas. We also introduce the two special states on the replicated Hilbert space and their corresponding graphical depiction
\begin{equation}
  \begin{aligned}
    \ket*{\circleSA}&=\smashoperator{\sum_{s_{1,2}\in \mathbb{Z}_q}}\ket{s_1,s_1,s_2,s_2}=
 \begin{tikzpicture}[baseline={([yshift=-0.6ex]current bounding box.center)},scale=0.5]
   \gridLine{0}{0}{0}{0.75}
   \circle{0}{0}
 \end{tikzpicture}\\
    \ket*{\squareSA}&=\smashoperator{\sum_{s_{1,2}\in \mathbb{Z}_q}}\ket{s_1,s_2,s_2,s_1}=
 \begin{tikzpicture}[baseline={([yshift=-0.6ex]current bounding box.center)},scale=0.5]
   \gridLine{0}{0}{0}{0.75}
   \square{0}{0}
 \end{tikzpicture}~.
  \end{aligned}
\end{equation}

Using the known expressions for the averages over the unitary group~\cite{collins2003moments} one can show that the averaged local gate obeys
\be
\begin{aligned}
\label{eq:local_gate_rels_1}
 &\begin{tikzpicture}[baseline={([yshift=-0.6ex]current bounding box.center)},scale=0.5]
   \prop{0}{0}{FcolU}{}
   \circle{0.5}{0.5}
    \circle{-0.5}{0.5}
 \end{tikzpicture}=
\begin{tikzpicture}[baseline={([yshift=-0.6ex]current bounding box.center)},scale=0.5]
    \gridLine{0}{0}{0}{0.75}
   \circle{0}{0.75}
    \gridLine{.5}{0}{.5}{0.75}
   \circle{.5}{0.75}
 \end{tikzpicture}~,& &
 \begin{tikzpicture}[baseline={([yshift=-0.6ex]current bounding box.center)},scale=0.5]
   \prop{0}{0}{FcolU}{}
   \square{0.5}{0.5}
    \square{-0.5}{0.5}
 \end{tikzpicture}=
 \begin{tikzpicture}[baseline={([yshift=-0.6ex]current bounding box.center)},scale=0.5]
    \gridLine{0}{0}{0}{0.75}
   \square{0}{0.75}
    \gridLine{.5}{0}{.5}{0.75}
   \square{.5}{0.75}
 \end{tikzpicture},\\
  &\begin{tikzpicture}[baseline={([yshift=-0.6ex]current bounding box.center)},scale=0.5]
   \prop{0}{0}{FcolU}{}
   \circle{0.5}{0.5}
    \square{-0.5}{0.5}
 \end{tikzpicture}=\alpha\left(
 \begin{tikzpicture}[baseline={([yshift=-0.6ex]current bounding box.center)},scale=0.5]
    \gridLine{0}{0}{0}{0.75}
   \circle{0}{0.75}
    \gridLine{.5}{0}{.5}{0.75}
   \circle{.5}{0.75}
 \end{tikzpicture}+ \begin{tikzpicture}[baseline={([yshift=-0.6ex]current bounding box.center)},scale=0.5]
    \gridLine{0}{0}{0}{0.75}
   \square{0}{0.75}
    \gridLine{.5}{0}{.5}{0.75}
   \square{.5}{0.75}
 \end{tikzpicture}\right)~,& &\begin{tikzpicture}[baseline={([yshift=-0.6ex]current bounding box.center)},scale=0.5]
   \prop{0}{0}{FcolU}{}
   \square{0.5}{0.5}
    \circle{-0.5}{0.5}
 \end{tikzpicture}=\alpha\left(
 \begin{tikzpicture}[baseline={([yshift=-0.6ex]current bounding box.center)},scale=0.5]
    \gridLine{0}{0}{0}{0.75}
   \circle{0}{0.75}
    \gridLine{.5}{0}{.5}{0.75}
   \circle{.5}{0.75}
 \end{tikzpicture}+ \begin{tikzpicture}[baseline={([yshift=-0.6ex]current bounding box.center)},scale=0.5]
    \gridLine{0}{0}{0}{0.75}
   \square{0}{0.75}
    \gridLine{.5}{0}{.5}{0.75}
   \square{.5}{0.75}
 \end{tikzpicture}\right),
\end{aligned} 
 \ee
 where $\alpha=q/(q^2+1)$. Similarly, one can establish the validity of the time-reversed expressions in which the diagrams are flipped about the horizontal axis. These expressions are exact for all $q$.  
 
In the folded representation we introduce two types of local initial states 
 \begin{align}
&\left(\ket{m_x}\otimes\ket{m_{x}}^\star\right)\otimes_{\rm r}\left(\ket{m_{x+y}}\otimes\ket{m_{x+y}}^\star\right)=\begin{tikzpicture}[baseline={([yshift=-0.6ex]current bounding box.center)},scale=0.65]
  \tgridLine{0}{0}{-0.25}{0.25}
  \tgridLine{1}{0}{1.25}{0.25}
  \istate{0}{0}{colSt}
  \node at (.5,-.625) {\scalebox{1}{$x$}};
   \node at (.5,.25) {\scalebox{1}{$y$}};
\end{tikzpicture}
 \end{align}
where $x$ refers to the position on the lattice, and $y\in\{0,1\}$ characterises the shift between the two copies. Setting $y=0$ gives the actual local initial state on the replicated space whereas $y=1$ implies a two-site shift between replicas.
Using these ingredients we can write~\footnote{{ Note that we chose the first argument equal to $-t-1$ so that the subsystem $A$ is positioned between sites $1$ and $2l$, cf.\ Eq.~\eqref{eq:D_def}.}}
\be
\expval{\smash{\tr[\smash{\rho_{A}^{(r)}(t)\rho_{A}^{(s)}(t)}]}}\equiv D_{|r-s|}(-t-1,2t+1,2\ell), 
\label{eq:overlapdiagram}
\ee 
where we introduced the diagram 
\begin{align}
\label{eq:D_def}
\!\!\!D_{y}(x,p,q)=\hspace{-1cm} \begin{tikzpicture}[baseline={([yshift=0.6ex]current bounding box.center)},scale=0.45]
    \foreach \x in {0,2,...,4}{\prop{\x}{0}{FcolU}{}}
    \foreach \x in {-1,1,...,5}{\prop{\x}{-1}{FcolU}{}}
    \foreach \x in {-2,0,...,6}{\prop{\x}{-2}{FcolU}{}}
    \foreach \x in {-3,-1,...,7}{\prop{\x}{-3}{FcolU}{}}
    \foreach \x in {-4,-2,...,8}{\prop{\x}{-4}{FcolU}{}}
    \foreach \x in {-5.5,-3.5,...,8.5}{\istate{\x}{-4.5}{colSt}}
    \foreach \x in {0,...,3}{\square{\x+0.5}{0.5}}
    \foreach \x in {-6,...,-1}{\circle{\x+0.5}{\x+1.5}}
    \foreach \x in {-6,...,-1}{\circle{4-\x-0.5}{\x+1.5}}
      \foreach \x in {-5,-3,...,9}{\node at (\x,-4.25) {\scalebox{0.7}{$y$}};}
    \node at (-5,-5.25) {\scalebox{0.7}{$x$}};
    \node at (-3,-5.25) {\scalebox{0.7}{$x+1$}};
    \node at (-1,-5.25) {\scalebox{0.7}{$\cdots$}};
    \node at (7,-5.25) {\scalebox{0.7}{$\cdots$}};
    \node at (9,-5.25) {\scalebox{0.7}{{$x\!+\!p\!+\!q\!-\!1$}}};
    \draw[semithick,decorate,decoration={brace}] (4.5,0.85) -- (9.85,-4.5) node[midway,xshift=7.5pt,yshift=7.5pt] {$p$};
    \draw[semithick,decorate,decoration={brace}] (0.25,0.75) -- (3.75,0.75) node[midway,above] {$2q$};
  \end{tikzpicture}
  \hspace{-.5cm},\hspace{-0.25cm}
\end{align}
which we have depicted for $p=6$ and $q=2$.

\begin{figure}
 \includegraphics[width=\columnwidth, trim= 35 20 175 0, clip]{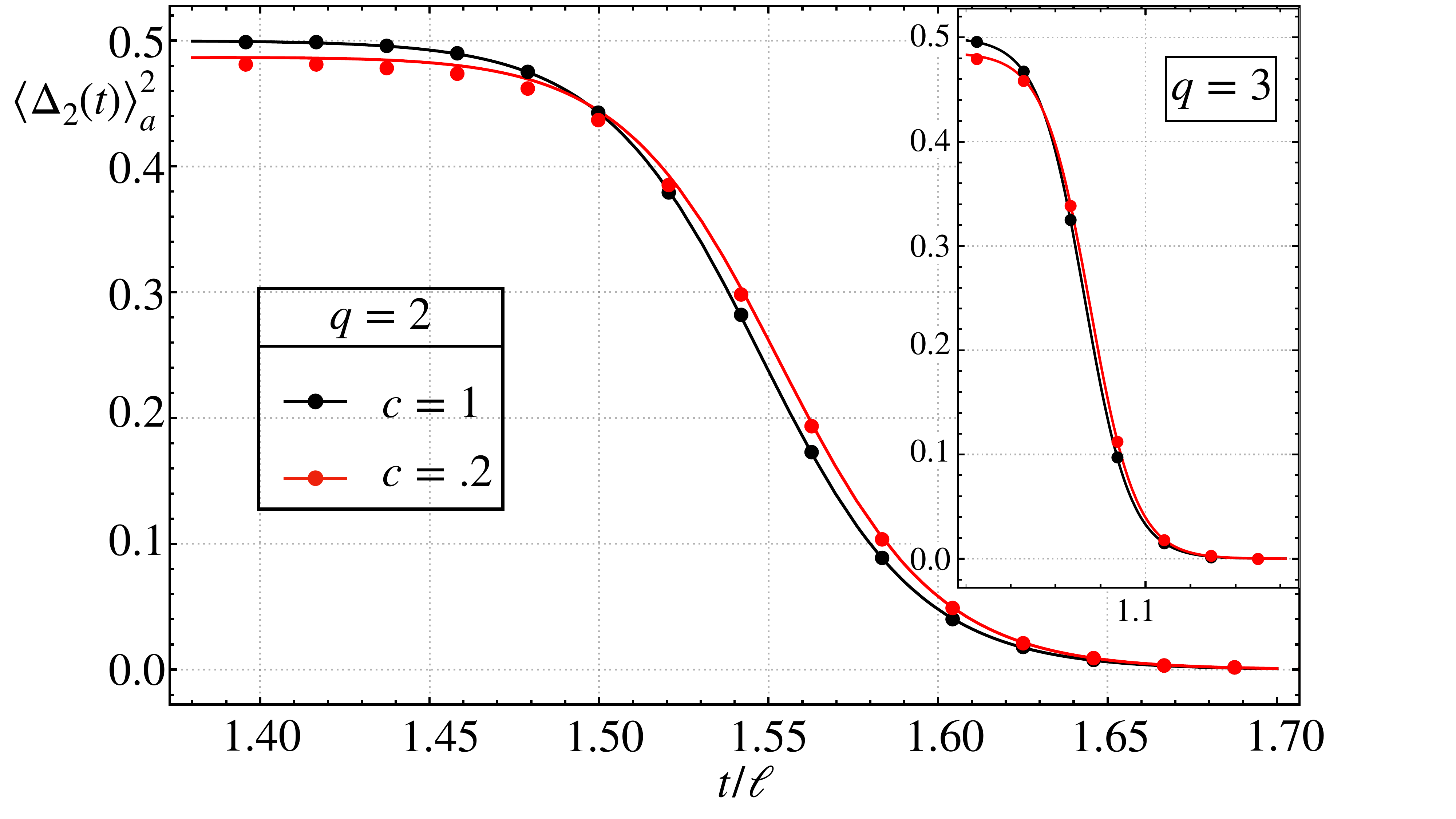}
  \caption{\label{fig:qme}
  The quantum Mpemba effect in random unitary circuits.  In the main figure we plot $\expval{\Delta_2(t)}_{\rm a}^2$ for two different states specified by~\eqref{eq:initialstate} using $c=1$ (black) and $c=.2$ (red)  with $q=2$ and $\ell=48$ as a function of $t/\ell$.  The symbols represent the exact solution of~\eqref{eq:recursion}, and~\eqref{eq:recursion_initial} while the solid lines use the analytic approximations~\eqref{eq:approximatesolution} (black) and an improved approximation (red) discussed in~\cite{Note11}.  We see that the lines cross indicating a weak quantum Mpemba effect, whereby the initially less symmetric state (black) $\epsilon$-approximately restores the symmetry faster. { For these parameters we have $v_\beta\approx 1.1$ (black) and $v_\beta\approx 0.68$ (red) while $v_E\approx0.64$. In the inset we plot the same quantity for $q=3$ showing a far less pronounced crossing. Here $v_\beta\approx 1.34$ (black), $v_\beta\approx0.96$ (red) while $v_E\approx 0.93$ }. }
\end{figure}
Employing the relations~\eqref{eq:local_gate_rels_1} to reduce the topmost line of gates in~\eqref{eq:D_def} we can relate $D_y(x,p,q)$ to a sum of 4 similar diagrams with $p-1$ rows but with different values of $x$ and $q$. Repeating this procedure we find that $D_y(x,p,q)$ satisfies the recurrence relation
\begin{align}
\!\!\!\!\!\!\!\!\!\! D_y(x,p,q)&=
 \alpha^2\left[
  D_{y}(x,p\!-\!1,q\!+\!1)\!+\!
  D_{y}(x,p\!-\!1,q)\right.\label{eq:recursion}\\
&+\!\!  \left.
  D_{y}(x\!+\!1,p\!-\!1,q)\!+\!
  D_{y}(x\!+\!1,p\!-\!1,q\!-\!1)
  \right],\nonumber
\end{align}
supplemented by the initial conditions
\begin{equation}\label{eq:recursion_initial}
  \begin{aligned}
    D_y(x,1,q)&=\gamma_y(x)\gamma_y(x+q)\prod_{j=1}^{q-1}\beta_y(x+j),\\
    D_y(x,p,0)&=1,
  \end{aligned}
\end{equation}
where we introduced 
\be
\begin{aligned}
\label{eq:beta}
  &\beta_y(x)\!=\!\tr [m_x m_{x+y}^\dag]\!\tr[m_{x+y}m_{x}^\dag]\equiv \begin{tikzpicture}[baseline={([yshift=-0.6ex]current bounding box.center)},scale=0.65]
  \istate{0}{0}{colSt}
  \node at (.5,-.625) {\scalebox{1}{$x$}};
   \node at (.5,.5) {\scalebox{1}{$y$}};
   \square{0}{0}
   \square{1}{0}
\end{tikzpicture},
\\
&\gamma_y(x)=\tr[m_xm_x^\dag m_{x+y}m_{x+y}^\dag]
  \equiv \begin{tikzpicture}[baseline={([yshift=-0.6ex]current bounding box.center)},scale=0.65]
  \istate{0}{0}{colSt}
  \node at (.5,-.625) {\scalebox{1}{$x$}};
   \node at (.5,.5) {\scalebox{1}{$y$}};
   \circle{0}{0}
   \square{1}{0}\end{tikzpicture}=\begin{tikzpicture}[baseline={([yshift=-0.6ex]current bounding box.center)},scale=0.65]
  \istate{0}{0}{colSt}
  \node at (.5,-.625) {\scalebox{1}{$x$}};
   \node at (.5,.5) {\scalebox{1}{$y$}};
   \square{0}{0}
   \circle{1}{0}\end{tikzpicture}.
\end{aligned}
\ee
Note that $\beta_y, \gamma_y(x)\in[0,1]$ and   $\beta_0=1$~\cite{Note111}.  The recursion relation \eqref{eq:recursion} can be solved exactly for arbitrary parameters.  The full solution is somewhat bulky and is presented~\cite{Note11} however, it takes a particularly simple and instructive form in the ``short-time'' regime $p\leq q$. In particular, plugging back into~\eqref{eq:overlapdiagram} we find  
\begin{equation}
\label{eq:shorttime}
\expval{\tr[\smash{\rho_{A}^{(r)}\rho_{A}^{(s)}}]}  
  = c_{r-s} \beta_{r-s}^{\ell-2t} ((1+\beta_{r-s})\alpha)^{4t},
\end{equation}
where the difference $r\!-\!s$ is computed mod 2 and $c_y$ is an initial-state dependent constant. We see that for $r=s$ this expression reproduces the known result for the averaged purity in random unitary circuits~\cite{nahum2018operator}. For $s\neq r$, the expression changes in two ways. First it acquires an exponential decay in the portion of $A$ that has not yet received information about $\bar A$, i.e., where the state is locally still pure. Second, its decay in time is dressed --- sped up --- by the initial-state dependent parameter $\beta_1$. 

 \footnotetext[111]{The last diagram of \eqref{eq:beta} is equal to $\mathrm{tr}[m^\dag_{x+y}m_{x+y}m^\dag_xm_x]$ which in general differs from $\gamma_y(x)$.  We consider states for which the two are the same. }

In fact, as illustrated in Fig.~\ref{fig:twistedstate}, for large $\ell$, $t$ the overlap is well approximated by combining short time expression and eventual relaxed value in a continuous manner, i.e.   
\begin{equation} \label{eq:approximatesolution1}
  \mkern-10mu
\expval{\smash{\tr[\smash{\rho_{A}^{(r)}\!\rho_{A}^{(s)}}]}}  
  \!\simeq\! \begin{cases}
    \! c_{r-s} \beta_{r-s}^{\ell-2t} ((1\!\!+\!\!\beta_{r-s})\alpha)^{4t}\!, 
    &t \leq \!{\ell}/{v_{\beta_{r-\!s}}} \\
    \!{q^{-2\ell}}, &t \geq \!{\ell}/{v_{\beta_{r-\!s}}}
\end{cases}
  \mkern-10mu
\end{equation}
where we introduced the $\beta$-dependent relaxation velocity 
\begin{equation}
  v_\beta= -2 \frac{\log(\alpha (\beta^{1/2}+\beta^{-1/2}))}{\log(q \beta^{1/2})}\in [v_E, 2],
\end{equation}
and the entanglement velocity $v_E= -2 {\log(2\alpha)}/{\log(q)}$. Plugging~\eqref{eq:approximatesolution1} into~\eqref{eq:delta2overlap} we find 
\be
\mkern-12mu
\expval{\Delta_2(t)}_{\rm a}^2 \simeq  \frac{1}{2}
\begin{cases}
  \displaystyle 1 - \frac{c_1}{c_0}  \beta_1^{\ell-2t} 
  \!\left(\!\frac{1+\beta_1}{2}\!\right)^{4t}\mkern-16mu,
  \mkern32mu
  &t \leq \frac{\ell}{v_{\beta_1}}, \\
\\[-1.75ex]
  \displaystyle 1 - \frac{1}{c_0}\frac{1}{q^{2\ell}(2\alpha)^{4t}}, 
  &\mkern-50mu
  \frac{\ell}{v_{\beta_1}} \leq t \leq \frac{\ell}{v_E},\\
\\[-1.75ex]
  \displaystyle 0,\, &t \geq \frac{\ell}{v_E}.\\
\end{cases}
\mkern-12mu
\label{eq:approximatesolution}
\ee 
We see that, although Eq.~\eqref{eq:approximatesolution} {displays} some explicit initial-state dependence, the latter is almost entirely washed away for large enough subsystems: for fixed $t/\ell < 1/v_E$ and large $\ell$,  Eq.~\eqref{eq:approximatesolution} approaches $1/2$ up to exponentially small corrections. Nevertheless, we see that for $\beta_1<1$ translation symmetry is approximately restored only for times {$t\simeq  \ell/v_E$}. More precisely, if for a given $\epsilon>0$, we ask what is the first time at which $\expval{\Delta_2(t)}_{\rm a}\leq \epsilon$ we find {$t \gtrsim \ell/v_E $}. This is also the time at which averaged Frobenius distance of $\rho_A(t)$ and $\1/q^{2\ell}$ approaches $\epsilon$. This means that the reduced state restores translation symmetry only when it thermalises and furthermore, it does so abruptly rather than gradually. 
{ As discusssed above, this appears to be in tension with the entanglement spectrum dynamics.  To alleviate this we note that $\Delta_2(t)$ encodes further information about the full matrix $\rho_A(t)$ beyond merely its spectrum,  meaning that some initial-state dependence remains encoded in its matrix structure.} Finally, note that Eq.~\eqref{eq:approximatesolution} vanishes identically for $\beta_1=1$, showing that the limit of the symmetric state commutes with that of large $\ell$ and $t$.  

These features are seen in Fig.~\ref{fig:twistedstate} where we plot $\expval{\Delta_2(t)^2}_{\rm a}$ for the state parametrized as
\begin{equation}\label{eq:initialstate}
    m_{2x-1}=\frac{\1+\sigma^z}{2}
    ,\qquad
    m_{2x}=\frac{\1+\sigma^z+c\sigma^x}{\sqrt{4+2 c^2}},
\end{equation}
where $\sigma^{x,z}$ are Pauli matrices acting on a 2 dimensional subspace of $\mathbb{C}^q$ and $c\in\mathbb R$. We see that as $\ell$ increases, $\expval{\Delta_2(t)}_{\rm a}$ behaves more like a step function and details of the state are forgotten. Likewise, for increasing $q$ the symmetry restoration is increasingly sharp. The case of $\nu>2$ can be analysed following the same logic and shows similar qualitative features~\cite{Note11}. 
 
For finite size subsystems, memory of the initial state is retained until large times.  This is exemplified in Fig.~\ref{fig:qme} where we again plot $\expval{\Delta_2(t)^2}_{\rm a}$ but 
for two choices of the parameter $c$ and Hilbert-space dimension $q=2,3$. We see that the initially more asymmetric state (black lines and symbols) exhibits $\epsilon$-approximate symmetry restoration \emph{before} the more symmetric state (red lines and symbols) signifying the occurrence of the quantum Mpemba effect.  The effect is less prominent for larger $q$ and vanishes for $q\to\infty$,  { which can be interpeted as the vanishing of the quantum Mpemba effect in the semi-classical limit.}

To summarise, in this work we have studied translation symmetry restoration in random unitary circuits. After introducing an observable-independent framework to conduct our analysis we showed that, { in contrast to what is suggested by the entanglement dynamics}, the restoration of translation symmetry only happens on time-scales proportional to the volume of the subsystem. Moreover, we have shown that for large enough subsystems the symmetry restoration process only depends on the degree of symmetry breaking introduced by the initial state, i.e.\  the number of two-site shifts under which it is invariant, no other detail of the state is retained. At intermediate sizes the initial-state dependence is still visible and one can observe instances of the quantum Mpemba effect which, nevertheless, vanishes in both the semiclassical and scaling limits. 

Our results on the approximate symmetry restoration are in agreement with the fact that random unitary circuits form approximate unitary $k$-designs for times linear in the system size~\cite{harrow2009random, diniz2011comment, brandao2013exponential, znidaric2008exact, brown2010convergence, brandao2016local, brandao2016efficient, hunterjones2019unitary, haferkamp2022randomquantum}. In essence, this means that when evolving up to those times, random unitary circuits become indistinguishable from all-to-all random unitary matrices acting on the full chain (and therefore all information about the initial state is lost). Our setting differs from that studied in this context in that we work in the thermodynamic limit. However, we find that a finite subsystem looses memory of the initial state only in a time that is linear in the size of the subsystem. 

Our work raises many interesting questions. An immediate one is whether more realistic systems, e.g.\ Hamiltonian systems without noise, also show simultaneous occurrence of spatial symmetry restoration and thermalisation. Our results suggest this to be the case: realistic systems should restore more slowly than random unitary ones (they retain more information) but all the symmetries are restored upon thermalisation. Another interesting question concerns the quantum Mpemba effect, which has previously been observed in non-integrable only through numerical results on small systems~\cite{liu2024symmetry,turkeshi2024quantum}. By analytical means we find that the effect invariably vanishes when system sizes are large.  This is in contrast to integrable systems, raising the question of how the effect is stabilized therein and its origins in the present case.

\begin{acknowledgments}
B.\ B.\ thanks Molly Gibbins and Adam Gammon-Smith for collaboration on related topics. We are grateful to Toma\v{z} Prosen and Tianci Zhou for useful comments on the draft. We acknowledge financial support from the Royal Society through the University Research Fellowship No.\ 201101 (B.\ B.), the European Research Council under Consolidator Grant number 771536 ``NEMO'' (C.\ R.), and from The Leverhulme Trust through the Early Career Fellowship No.\ ECF-2022-324 (K.\ K.). 
\end{acknowledgments}

\bibliography{Random_Mpemba.bib}

\input{SM.tex}

\end{document}

%% file: SM.tex
\onecolumngrid
\newpage 
\newcounter{equationSM}
\newcounter{figureSM}
\newcounter{tableSM}
\stepcounter{equationSM}
\setcounter{equation}{0}
\setcounter{figure}{0}
\setcounter{table}{0}
\setcounter{section}{0}
\makeatletter
\renewcommand{\theequation}{\textsc{sm}-\arabic{equation}}
\renewcommand{\thefigure}{\textsc{sm}-\arabic{figure}}
\renewcommand{\thetable}{\textsc{sm}-\arabic{table}}

\begin{center}
  {\large{\bf Supplemental Material for ``Translation Symmetry Restoration under Random Unitary Dynamics''}}
\end{center}

Here we report some useful information complementing the main text. In particular
\begin{itemize}
   \item[-] In Sec.~\ref{sec:recursion} we provide the solution to the recursion relations for $D_y(x,m,p)$.
   \item[-] In Sec.~\ref{sec:generalization} we discuss the generalization to higher $\nu$.
\end{itemize}
 
\section{Exact solution of recursion relations for $D_y(x,m,p)$ }
\label{sec:recursion}
The generalized averaged purities admit a simple depiction in the graphical notation,
\be
\left<\tr[\rho_A^{(a)}(t)\rho_{A}^{(b)}(t)]\right>=\begin{tikzpicture}[baseline={([yshift=-0.6ex]current bounding box.center)},scale=0.5]
\foreach \x in {-9,-7,...,7}{\prop{\x}{0}{FcolU}{}}
   \foreach \x in {-8,-6,...,6}{\prop{\x}{-1}{FcolU}{}}
   \foreach \x in {-9,-7,...,7}{\prop{\x}{-2}{FcolU}{}}
   \foreach \x in {-8,-6,...,6}{\prop{\x}{-3}{FcolU}{}}
   \foreach \x in {-9,-7,...,7}{\prop{\x}{-4}{FcolU}{}}
   \foreach \x in {-8,-6,...,6}{\prop{\x}{-5}{FcolU}{}}
   \foreach \x in {-9.5,-7.5,...,7.5}{\istate{\x}{-5.5}{colSt}};
   \foreach \x in {-9.5,...,-4.5}{\circle{\x}{.5}}   
   \foreach \x in {7.5,...,2.5}{\circle{\x}{.5}}    \foreach \x in {-3.5,...,1.5}{\square{\x}{.5}}   
   \foreach \x in {-9,-7,...,7}{\node at (\x,-5.25) {\scalebox{0.7}{$b-a$}};}
   \node at (-5,-6.25) {\scalebox{0.7}{$L+a$}};
   \node at (-3,-6.25) {\scalebox{0.7}{$1+a$}};
   \node at (0,-6.25) {\scalebox{0.7}{$\dots\dots$}};
   \node at (3,-6.25) {\scalebox{0.7}{$\ell$}};
   \node at (5,-6.25) {\scalebox{0.7}{$\dots$}};
   \node at (-8,-6.25) {\scalebox{0.7}{$\dots$}};
 \end{tikzpicture},\qquad
 a\le b,\quad L+1\equiv 0,
\ee
where we have depicted the case of $t=3$ and $\ell=4$. Note that due to the cyclicity of the trace we can without loss of generality take $a\le b$.  
Using the properties of the local gates,  $ \begin{tikzpicture}[baseline={([yshift=-0.6ex]current bounding box.center)},scale=0.5]
  \prop{0}{0}{FcolU}{}
  \circle{0.5}{0.5}
   \circle{-0.5}{0.5}
\end{tikzpicture}=
\begin{tikzpicture}[baseline={([yshift=-0.6ex]current bounding box.center)},scale=0.5]
   \gridLine{0}{0}{0}{0.75}
  \circle{0}{0.75}
   \gridLine{.5}{0}{.5}{0.75}
  \circle{.5}{0.75}
\end{tikzpicture}~,
\begin{tikzpicture}[baseline={([yshift=-0.6ex]current bounding box.center)},scale=0.5]
  \prop{0}{0}{FcolU}{}
  \square{0.5}{0.5}
   \square{-0.5}{0.5}
\end{tikzpicture}=
\begin{tikzpicture}[baseline={([yshift=-0.6ex]current bounding box.center)},scale=0.5]
   \gridLine{0}{0}{0}{0.75}
  \square{0}{0.75}
   \gridLine{.5}{0}{.5}{0.75}
  \square{.5}{0.75}
\end{tikzpicture} $,
we can reduce this diagram to the quantity $D_{y}(-t-1,2t+1,\ell)$, defined as
\be
D_y(x,m,p)=
\begin{tikzpicture}[baseline={([yshift=-0.6ex]current bounding box.center)},scale=0.5]
   \foreach \x in {0,2,...,6}{\prop{\x}{0}{FcolU}{}}
   \foreach \x in {-1,1,...,7}{\prop{\x}{-1}{FcolU}{}}
   \foreach \x in {-2,0,...,8}{\prop{\x}{-2}{FcolU}{}}
   \foreach \x in {-3,-1,...,9}{\prop{\x}{-3}{FcolU}{}}
   \foreach \x in {-4,-2,...,10}{\prop{\x}{-4}{FcolU}{}}
   \foreach \x in {-5.5,-3.5,...,11}{\istate{\x}{-4.5}{colSt}}
   \foreach \x in {0,...,5}{\square{\x+0.5}{0.5}}
   \foreach \x in {-6,...,-1}{\circle{\x+0.5}{\x+1.5}}
   \foreach \x in {-6,...,-1}{\circle{6-\x-0.5}{\x+1.5}}
   \foreach \x in {-5,-3,...,11}{\node at (\x,-4.25) {\scalebox{0.7}{$y$}};}
   \node at (-5,-5.25) {\scalebox{0.7}{$x$}};
   \node at (-3,-5.25) {\scalebox{0.7}{$x+1$}};
   \node at (-1,-5.25) {\scalebox{0.7}{$x+2$}};
   \node at (1,-5.25) {\scalebox{0.7}{$x+3$}};
   \node at (3,-5.25) {\scalebox{0.7}{$\cdots$}};
   \node at (9,-5.25) {\scalebox{0.7}{$\cdots$}};
   \node at (11,-5.25) {\scalebox{0.7}{$x+m+p-1$}};
   \draw[semithick,decorate,decoration={brace}] (6.5,0.85) -- (11.85,-4.5) node[midway,xshift=7.5pt,yshift=7.5pt] {$m$};
   \draw[semithick,decorate,decoration={brace}] (0.25,0.75) -- (5.75,0.75) node[midway,above] {$2p$};
 \end{tikzpicture}.
\ee
From this we can then employ the relations $\begin{tikzpicture}[baseline={([yshift=-0.6ex]current bounding box.center)},scale=0.5]
  \prop{0}{0}{FcolU}{}
  \circle{0.5}{0.5}
   \square{-0.5}{0.5}
\end{tikzpicture}=\alpha\left(
\begin{tikzpicture}[baseline={([yshift=-0.6ex]current bounding box.center)},scale=0.5]
   \gridLine{0}{0}{0}{0.75}
  \circle{0}{0.75}
   \gridLine{.5}{0}{.5}{0.75}
  \circle{.5}{0.75}
\end{tikzpicture}+ \begin{tikzpicture}[baseline={([yshift=-0.6ex]current bounding box.center)},scale=0.5]
   \gridLine{0}{0}{0}{0.75}
  \square{0}{0.75}
   \gridLine{.5}{0}{.5}{0.75}
  \square{.5}{0.75}
\end{tikzpicture}\right)~, \begin{tikzpicture}[baseline={([yshift=-0.6ex]current bounding box.center)},scale=0.5]
  \prop{0}{0}{FcolU}{}
  \square{0.5}{0.5}
   \circle{-0.5}{0.5}
\end{tikzpicture}=\alpha\left(
\begin{tikzpicture}[baseline={([yshift=-0.6ex]current bounding box.center)},scale=0.5]
   \gridLine{0}{0}{0}{0.75}
  \circle{0}{0.75}
   \gridLine{.5}{0}{.5}{0.75}
  \circle{.5}{0.75}
\end{tikzpicture}+ \begin{tikzpicture}[baseline={([yshift=-0.6ex]current bounding box.center)},scale=0.5]
   \gridLine{0}{0}{0}{0.75}
  \square{0}{0.75}
   \gridLine{.5}{0}{.5}{0.75}
  \square{.5}{0.75}
\end{tikzpicture}\right)$ (cf.\ Eq.~\eqref{eq:local_gate_rels_1}) to remove the topmost row of local gates and obtain the following recurrence relation,
\be
 D_{y}(x,m,p)=
\alpha^2\left[ D_{y}(x,m-1,p+1)+ D_{y}(x,m-1,p)+ D_{y}(x+1,m-1,p)+ D_{y}(x+1,m-1,p-1) \right].
\ee
To be able to solve it, we need to fix two initial conditions. These are obtained from the $p=0$ and $m=1$ points
\begin{equation}
\begin{aligned}
  D_y(x,m,0)&=
 \begin{tikzpicture}[baseline={([yshift=-0.6ex]current bounding box.center)},scale=0.5]
   \foreach \x in {0}{\prop{\x}{0}{FcolU}{}}
   \foreach \x in {-1,1}{\prop{\x}{-1}{FcolU}{}}
   \foreach \x in {-2,0,2}{\prop{\x}{-2}{FcolU}{}}
   \foreach \x in {-3,-1,...,3}{\prop{\x}{-3}{FcolU}{}}
   \foreach \x in {-4,-2,...,4}{\prop{\x}{-4}{FcolU}{}}
   \foreach \x in {-5.5,-3.5,...,4.5}{\istate{\x}{-4.5}{colSt}}
   \foreach \x in {-6,...,-1}{\circle{\x+0.5}{\x+1.5}}
   \foreach \x in {-6,...,-1}{\circle{-\x-0.5}{\x+1.5}}
   \foreach \x in {-5,-3,...,5}{\node at (\x,-4.25) {\scalebox{0.7}{$y$}};}
   \node at (-5,-5.25) {\scalebox{0.7}{$x$}};
   \node at (-3,-5.25) {\scalebox{0.7}{$x+1$}};
   \node at (-1,-5.25) {\scalebox{0.7}{$x+2$}};
   \node at (1,-5.25) {\scalebox{0.7}{$\cdots$}};
   \node at (3,-5.25) {\scalebox{0.7}{$\cdots$}};
   \node at (5,-5.25) {\scalebox{0.7}{$x+m-1$}};
   \draw[semithick,decorate,decoration={brace}] (.5,0.85) -- (5.85,-4.5) node[midway,xshift=7.5pt,yshift=7.5pt] {$m$};
 \end{tikzpicture}=1,\\
  D_y(x,1,p)&=\begin{tikzpicture}[baseline={([yshift=-0.6ex]current bounding box.center)},scale=0.5]
   \foreach \x in {-5.5,-3.5,...,5}{\istate{\x}{-4.5}{colSt}}
   \foreach \x in {-5,...,4}{\square{\x+0.5}{-6+1.5}}
   \foreach \x in {-6,5}{\circle{\x+0.5}{-6+1.5}}
   \foreach \x in {-5,-3,...,5}{\node at (\x,-4.25) {\scalebox{0.7}{$y$}};}
   \node at (-5,-5.25) {\scalebox{0.7}{$x$}};
   \node at (-3,-5.25) {\scalebox{0.7}{$x+1$}};
   \node at (1,-5.25) {\scalebox{0.7}{$\cdots$}};
   \node at (5,-5.25) {\scalebox{0.7}{$x+p-1$}};
   \draw[semithick,decorate,decoration={brace}] (-4.5,0.75-4.5) -- (4.5,0.75-4.5) node[midway,above] {$2p$};
 \end{tikzpicture}=\gamma^L_y(x)\prod _{j=1}^{p-1}\beta_{y}(x+j)\gamma^R_y(x+p),
\end{aligned}
\end{equation}
with the coefficients $\beta_{y}(x)$, $\gamma^{L/R}_{y}(x)$ defined in Eq.~\eqref{eq:beta}.
Note that in the main text we assumed $\gamma^{L}_y(x)=\gamma^{R}_y(x)\equiv\gamma_{y}(x)$, but this is only true for the initial states that we consider. In general the two coefficients are different,
\begin{equation}
  \gamma^{L}_y(x)=\tr[m_xm_x^\dag m_{x+y}m_{x+y}^\dag],\qquad
  \gamma^{R}_y(x)=\tr[m_x^{\dag}m_x m_{x+y}^{\dagger}m_{x+y}].
\end{equation}
The full solution to this set of equations can be straightforwardly shown to be given by
\begin{equation}
 \begin{aligned}
   D_{y}(x,m,p)=
   \alpha^{2(m-1)}
   \Bigg[&\sum_{k_1=0}^{m-1}\sum_{k_2=\mathrm{max}\{0,k_1-p+1\}}^{m-1}
     \binom{m-1}{k_1}\binom{m-1}{k_2}
     \gamma^L_y(x+k_1)
     \smashoperator{\prod _{j=1}^{p+k_2-k_1-1}}\beta_{y}(x+k_1+j)\gamma^R_y(x+p+k_2)  \\
     &-\sum_{k_1=p}^{m-2}\sum_{k_2=k_1+1-p}^{m-p-1}
     \binom{m-1}{k_1-p}\binom{m-1}{k_2+p}
     \gamma^L_y(x+k_1)\smashoperator{\prod _{j=1}^{p+k_2-k_1-1}}
     \beta_{y}(x+k_1+j)\gamma^R_y(x+p+k_2) \\
     & + \sum _{n=p}^{m-1} \frac{1}{\alpha^{2(m-n-1)}}\left(\binom{2 n-1}{n-p}-\binom{2 n-1}{n+p}\right)+\frac{\delta _{p,0}}{{\alpha^{2(m-1)}}} \Bigg].
 \end{aligned}
 \label{eq:fullsolution}
\end{equation}

Whenever $m\le p$ only the first term in the above expression gives a non-zero contribution and we have
\begin{equation}\label{eq:fullEarlyTime}
   D_{y}(x,m,p)=
   \alpha^{2(m-1)}
   \sum_{k_1=0}^{m-1}\sum_{k_2=0}^{m-1}
     \binom{m-1}{k_1}\binom{m-1}{k_2}
     \gamma^L_y(x+k_1)
     \smashoperator{\prod _{j=1}^{p+k_2-k_1-1}}\beta_{y}(x+k_1+j)\gamma^R_y(x+p+k_2),
\end{equation}
which can be further simplified by specializing the coefficients $\beta_{y}(x)$, $\gamma^{L/R}_y(x)$ to our case. In particular, we note that we can reparametrise them as
\begin{equation}
\beta_0(x)=1,\quad \beta_1(x)=\beta_1(x+1)=:\beta_1,\qquad
  \gamma^{L/R}_1(x)=\gamma^{L/R}_1(x)=:\gamma_1^{L/R},\quad
  \gamma^{L}_0(x)=\gamma_0^{R}(x)=:\gamma_0^{\mathrm{s}}+(-1)^x \gamma_0^{\mathrm{a}},
\end{equation}
which inserted into~\eqref{eq:fullEarlyTime} gives
\begin{equation}
  D_0(x,m,p)= \alpha^{2(m-1)} 2^{2(m-1)}\left.\gamma_0^{\mathrm{s}}\right.^2,\qquad
  D_{1}(x,m,p)=\gamma^{L}_1\gamma^{R}_1\beta^{p-m}\alpha^{2(m-1)}(1+\beta)^{2(m-1)},
\end{equation}
which gives Eq.~\eqref{eq:shorttime}. Doing the same substitution directly in~\eqref{eq:fullsolution} and considering an appropriate scaling limit for $m>p$ one can recover Eq.~\eqref{eq:approximatesolution1}.

\section{Generalization to  $\nu>2$}
\label{sec:generalization}

The majority of the analysis of the main text centered on states with $\nu=2$ i.e. those which intially broke translational symmetry by 4 sites.  The extension to $\nu>2$,  however, can be carried out straighforwardly with a similar phenomenology emerging. To do this, we extend the notation of the main text and write 
\begin{eqnarray}
\rho^{(k)}_A(t)=\tr_{\bar A}\left[\Pi^k\ketbra{\Psi(t)}{\Psi(t)}\Pi^k\right]
\end{eqnarray}
where again $\Pi$ is the two-site shift operator, $\nu>k\in \mathbb{N}_0$ and $\Pi^\nu\ket{\Psi(t)}=\ket{\Psi(t)}$ for states invariant under $2\nu$-site shifts.  In terms of this we have that 
\begin{eqnarray}\label{eq:DeltaGeneral}
\expval{\Delta_2(t)}^2_{\rm a}=1-\frac{2}{\nu}\sum_{k=0}^{\nu-1}\frac{\tr [\rho^{(0)}_A(t)\rho^{(k)}_A(t)]}{\tr [\left(\rho^{(0)}_A(t)\right)^2]}+\frac{1}{\nu^2}\sum_{j,k=0}^{\nu-1}\frac{\tr [\rho^{(j)}_A(t)\rho^{(k)}_A(t)]}{\tr [\left(\rho^{(0)}_A(t)\right)^2]}.
\end{eqnarray}
Here we see that the higher $\nu$ case is a sum over terms which resemble the $\nu=2$ case and so the same reasoning can be applied. In particular, both the numerator and denominator in  each of the terms, apart from the first, will take the form of~\eqref{eq:approximatesolution1}. Accordingly, the full sum results in an approximate form similar to~\eqref{eq:approximatesolution} and hence the symmetry will remain broken up to times which are linear in the subsystem size.  The maximal value can be shown to be $\frac{\nu-1}{\nu}$ which occurs when all shifted states are orthogonal to each other and the original state at $t=0$.

\begin{figure}
\includegraphics[width=0.48\columnwidth, trim= 20 0 80 0, clip]{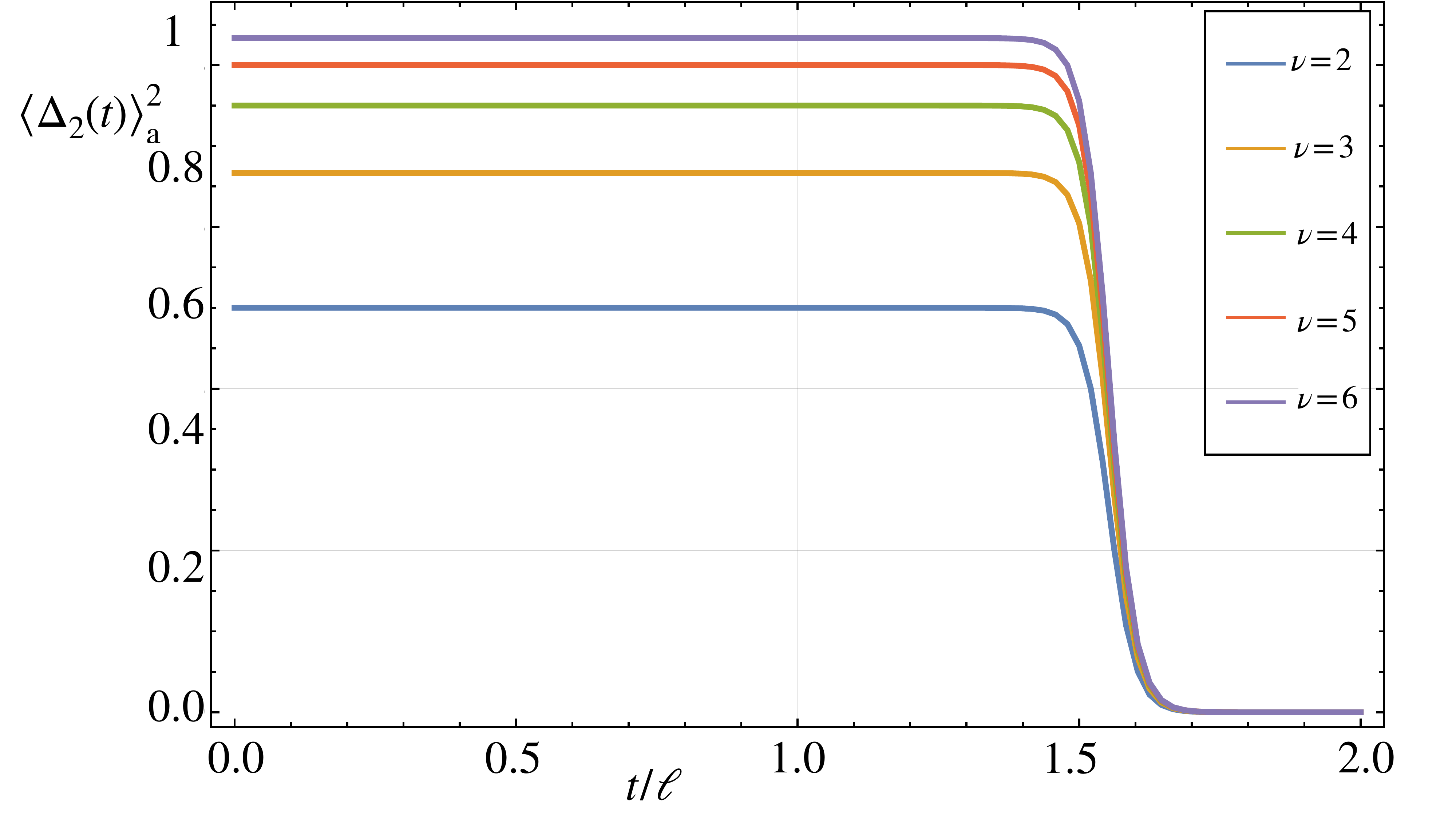}
\includegraphics[width=0.48\columnwidth, trim= 20 0 80 0, clip]{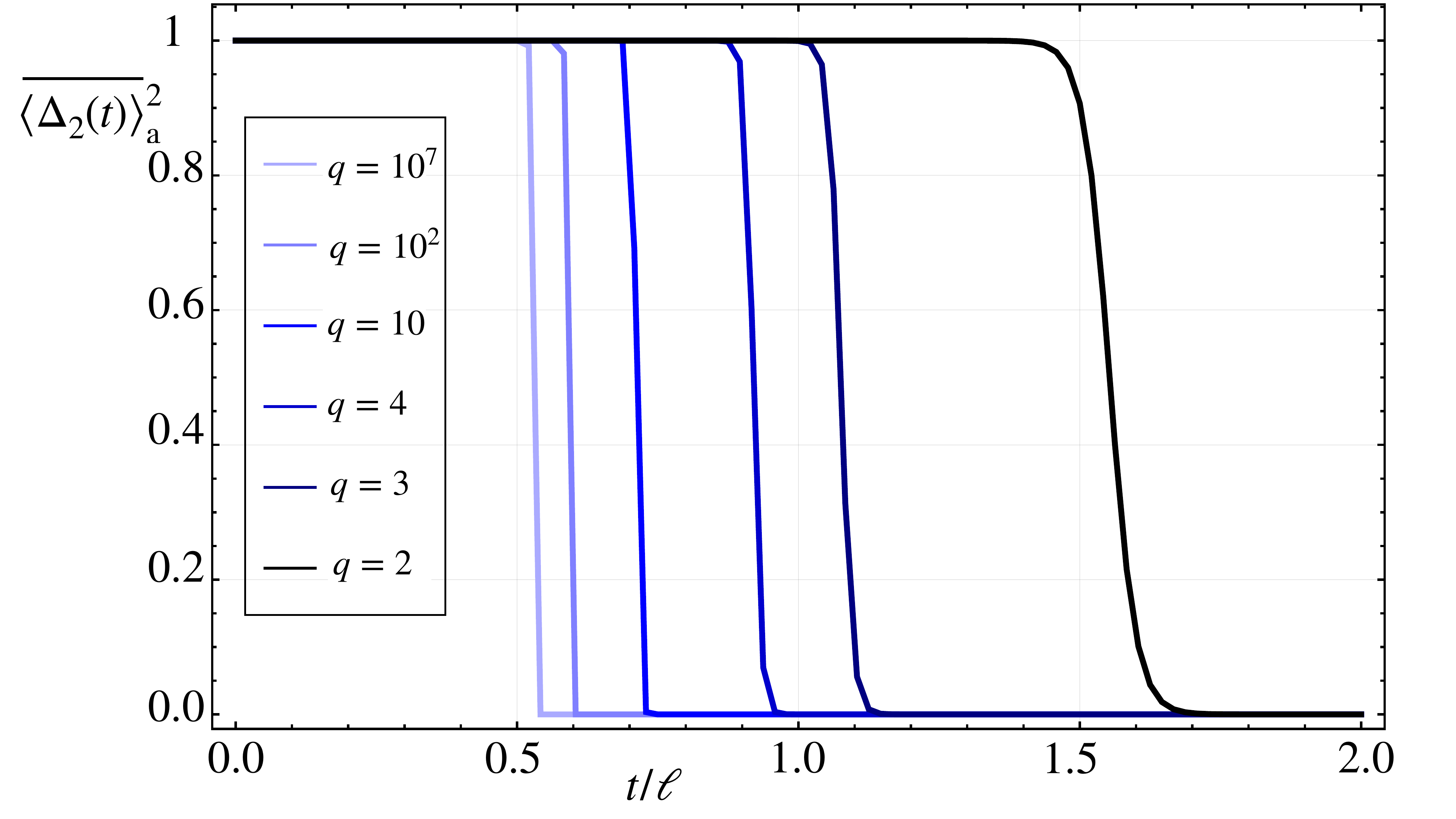}
  \caption{\label{fig:Supptwistedstate}
 Dynamical restoration of $2$-site translational symmetry from a $2\nu$-site spin helix.  On the left we plot $\expval{\Delta_2(t)}_{\rm a}^2$ for different values $\nu=2,\dots,6$ as a function of $t/\ell$ for $q=2,\ell=48$.  After rescaling by $\frac{\nu}{\nu-1}$ all lines collapse on to each other.  This is shown on the right where we plot $\overline{\expval{\Delta_2(t)}_{\rm a}^2}\equiv \frac{\nu}{\nu-1}\expval{\Delta_2(t)}_{\rm a}^2$ for $\ell=48$ for increasing $q$,darker to lighter. We see that symmetry resotration occurs earlier for larger $q$ and approaches a Heaviside function about $t=\ell/2$ for $q\to\infty$. }
\end{figure}

To illustrate these ideas,  we consider a particular initial state -- a spin helix --  parameterized by  
\begin{equation}
  m_j=\frac{1}{2}\left(\1
  +\cos{\left(\frac{2\pi j}{\nu}\right)}\sigma^z
  +\sin{\left(\frac{2\pi j}{\nu}\right)}\sigma^x\right),
\end{equation}
where $\sigma^{x,z}$ are Pauli matrices acting on a 2 dimensional subspace of $\mathbb{C}^q$.  In the case of $q=2$ this state is comprised of  consecutive two-site blocks of spins rotated about the $y$-axis by an angle $\pi/\nu$ relative to the previous block, starting from both spins in pointing in the $z$-direction.  For higher $q$  the same description applies with the two spin states being any two choices of orthogonal vectors in $\mathbb{C}^q$ and the $\sigma^x,\sigma^z$ acting as normal on the basis of these two states.  For this choice we can evaluate the terms of the sum~\eqref{eq:DeltaGeneral} above and find 
\begin{eqnarray}
\tr [\rho^{(j)}_A(t)\rho^{(k)}_A(t)]=\tr [\rho^{(0)}_A(t)\rho^{(|j-k|)}_A(t)]
\end{eqnarray} 
which can be computed using 
\begin{equation}
\beta_{|j-k|}=\cos^4\left({|j-k|\pi}/{\nu}\right),~\gamma_{|j-k|}=\cos^2\left({|j-k|\pi}/{\nu}\right).
\end{equation}
In Fig~\ref{fig:Supptwistedstate} we plot $\expval{\Delta_2(t)}^2_{\rm a}$ for the random unitary dynamics emerging from the spin helix state for $\nu=2,\dots,6$ as a function of $t/\ell$.  On the left we see that as in the case of $\nu=2$ the symmetry remains broken up till a time which is linear in subsystem size.  In fact, for this choice, the symmetry restoration is only trivially dependent on $\nu$. Specifically, we find that
\begin{equation}
  \left.\expval{\Delta_2(t)}_{\rm a}^2\right|_{\nu>2}
  =2\frac{\nu-1}{\nu}\left.\expval{\Delta_2(t)}_{\rm a}^2\right|_{\nu=2},
\end{equation}
which in particular means that the symmetry restoration is independent of the extent of the symmetry breaking.  On the right of Fig~\ref{fig:Supptwistedstate} we plot the rescaled Frobenious distance 
\be
\overline{\expval{\Delta_2(t)}_{\rm a}^2}=\frac{\nu}{\nu-1}\expval{\Delta_2(t)}_{\rm a}^2
\ee
for the same values showing all curves collapse onto one.  In addition, we plot the same quantity for increasing values of $q=2,\dots,10^7$. We see the same phenomenology emerge, namely; for larger local Hilbert space dimension there is faster symmetry restoration and moreover
\begin{eqnarray}
\lim_{q\to\infty}\expval{\Delta_2(t)}_{\rm a}^2=\frac{\nu-1}{\nu}\Theta(\ell-2t).
\end{eqnarray}

